\documentclass[english,pra,twocolumn,showpacs,amsmath,amssymb,lengthcheck]{revtex4-1}

\usepackage[normalem]{ulem}
\usepackage{hyperref}
\usepackage{graphicx}
\usepackage[]{color}

\def\be{\begin{equation}}
\def\ee{\end{equation}} 
\def\bea{\begin{eqnarray}}
\def\eea{\end{eqnarray}} 
\def\ba{\begin{array}} 
\def\ea{\end{array}}

\def\eps{\varepsilon}

\def\om{\omega}

\def\nn{\nonumber}
\def\ket{\rangle}
\def\bra{\langle}
\def\b{\mathbf}
\def\lra{\leftrightarrow}
\def\f{\frac}

\newcommand{\comments}[1]{}

\begin{document}
\title{Finite-momentum Bose-Einstein condensates in shaken 2D square optical lattices}
\author{M. Di Liberto$^{1,2}$}
\author{O. Tieleman$^1$}
\author{V. Branchina$^{3,4}$}
\author{C. Morais Smith$^1$}
\affiliation{$^1$Institute for Theoretical Physics, Utrecht University, Leuvenlaan 4, 3584CE Utrecht, the Netherlands}
\affiliation{$^2$Scuola Superiore di Catania, Universit\`{a} di Catania, Via Valdisavoia 9, I-95123 Catania, Italy}
\affiliation{$^3$Department of Physics, University of Catania}
\affiliation{$^4$INFN, Sezione di Catania, Via Santa Sofia 64, I-95123 Catania, Italy}
\date{\today}

\begin{abstract}
We consider ultracold bosons in a 2D square optical lattice described by the Bose-Hubbard model. In addition, an external time-dependent sinusoidal force is applied to the system, which shakes the lattice along one of the diagonals. The effect of the shaking is to renormalize the nearest-neighbor hopping coefficients, which can be arbitrarily reduced, can vanish, or can even change sign, depending on the shaking parameter. It is therefore necessary to account for higher-order hopping terms, which are renormalized differently by the shaking, and introduce anisotropy into the problem. We show that the competition between these different hopping terms leads to finite-momentum condensates, with a momentum that may be tuned via the strength of the shaking. We calculate the boundaries between the Mott-insulator and the different superfluid phases, and present the time-of-flight images expected to be observed experimentally. Our results open up new possibilities for the realization of bosonic analogs of the FFLO phase describing inhomogeneous  superconductivity. 
\end{abstract}


\maketitle

\section{INTRODUCTION}

Ultracold atoms in optical lattices are ideal systems to simulate the quantum behavior of condensed matter because the lattice geometry, the type of atoms (bosons or fermions), and their interactions can be manipulated in a perfectly clean environment \cite{bloch}. Furthermore, they provide a perfect testing ground for a wide variety of theoretical models. One of the most prominent examples is the Bose-Hubbard model, which has been studied extensively theoretically (e.g.~Refs.~\cite{fisher, jaksch}) and realised experimentally \cite{greiner, spielman}. 

More recently, much interest has been devoted to time-dependent, periodically stirred optical lattices, which allow for engineering synthetic gauge fields into the system \cite{lincoln,williams,gemelke}. 
In the presence of a staggered rotation, Dirac cones were shown to emerge for square optical lattices, thus simulating the behavior of anisotropic graphene when the system is loaded with fermions \cite{lim2}. If instead a mixture of fermions and bosons is used, several properties of high-$T_c$ superconductors can be reproduced \cite{lim4,lim3}. Loading the same lattice with dipolar bosons leads to a supersolid phase with vortices \cite{olivier}. 

Besides the interesting features that arise in the presence of rotation, a full range of new possibilities was shown to emerge by shaking the optical lattice. If the shaking frequency is much larger than the other characteristic energy scales in the problem, the parameters of the Hamiltonian are renormalized. This provides another tool to control the lattice parameters, and even enables the simulation of otherwise experimentally inaccessible lattice models \cite{eckardt,eckardt1,hemmerich}.
In the Bose-Hubbard model, for instance, the superfluid-Mott-insulator transition has been driven by ramping the shaking perturbation and thus tuning the effective hopping parameter to zero \cite{zenesini,*lignier}. Another fascinating experiment has revealed that magnetically frustrated systems can be realized with spinless bosons by applying elliptical shaking to a triangular lattice \cite{sengstock}.

Here, we consider a 2D square lattice shaken along one diagonal and investigate the effect of next-nearest-neighbor (nnn) and next-next-nearest neighbor (nnnn) hopping in the behavior of a bosonic system. In an effective description, the shaking perturbation leads to a renormalization of the nearest-neighbor (nn)  hopping parameter, which can vanish or even become negative \cite{eckardt,hemmerich}. When this parameter is tuned to be very small, higher order hopping terms, which are usually negligible, may become relevant and must therefore be included in the model. Although the nnn hopping coefficients are strictly zero in 2D optical lattices where the $x-$ and $y-$ directions are independent (separable potential), they are relevant for  non-separable optical lattices.  In this paper, we show that a {\it tunable} finite-momentum  condensate can be realized in a certain range of parameters for a realistic and simple setup, thus bringing us a step further in the  realization and control of finite-momentum Bose-Einstein condensates (BECs). 

Finite-momentum condensates have recently attracted a great deal of attention. In the original proposals by Fulde, Ferrel, Larkin, and Ovchinnikov (FFLO), it was argued that finite momentum Cooper pairs would lead to inhomogeneous superconductivity, with the superconducting order parameter varying spatially (the so-called FFLO phase) \cite{fulde,*larkin}. Early NMR experiments at high magnetic fields and low temperatures in the heavy-fermion compound CeCoIn$_5$ have shown indications of an FFLO phase \cite{bianchi,*radovan,*kenzelmann}, although recent data suggest the existence of a more complex phase, where the exotic FFLO superconductivity coexists with an incommensurate spin-density wave \cite{kumagai}. For ultracold fermions with spin imbalance, on the other hand, the observation of the FFLO phase has been recently reported in 1D \cite{liao}.  

Earlier theoretical studies of a square-lattice toy model for a scalar field, which took into account non-trivial hopping beyond nearest neighbors, have shown that quantum phases may be generated in which the order parameter is modulated in space \cite{branchina}. Finite-momentum condensates were also experimentally detected for bosons in more complex lattice  geometries, such as the triangular lattice under elliptical shaking \cite{sengstock}, or for more complex interactions, as e.g. for spinor  bosons in  a trap in the presence of Zeeman and spin-orbit interactions \cite{lin2}. With regard to bosons in a square lattice, it was recently shown that a staggered gauge field may lead to finite-momentum condensates \cite{lim}. In this case, the bosons condense either at zero momentum or in the corner of the Brillouin zone, and a first-order phase transition occurs between these two phases \cite{lim}. Here, we propose that finite-momentum condensates can be realized for bosons in a shaken square lattice and that we may {\it tune} the momentum of the condensate smoothly from $\b{0}$ to $(\pi,\pi)$, by varying the shaking parameter $K_0$. To the best of our knowledge, this is the first time that such an effect has been predicted for optical lattices as originating solely from beyond-nearest-hopping terms. The interaction simply shifts the ground-state energy by a constant and does not change the condensation momentum.

In the following, we consider in Sec.\ II an extended Bose-Hubbard model which includes higher-order-hopping coefficients for a non-separable 2D square optical lattice.  After introducing a sinusoidal shaking force to the system, we show in Sec.\ III how the finite-momentum condensate arises, and how the condensation momentum depends on the shaking. 
We present a 3D phase diagram, with as parameters the Hubbard interaction $U$, the chemical potential $\mu$, and the shaking parameter $K_0$, and indicate the required parameters for the realization of the tunable regime in Sec.\ IV. Finally, we calculate the expected outcome of time-of-flight experiments in Sec.\ V and present our conclusions in Sec.\ VI. 

\section{THE MODEL} 

Before discussing the generic 2D problem, let us recall the behavior of 1D lattices and 2D separable lattices. A simple calculation shows that in 1D optical lattices of the form $V(x)=(V_0/2)\cos(2\tilde{k}x)$ ($V_0$ is the potential depth and $\tilde{k}=2\pi/\lambda$ is the wave vector of the laser beam), nnn hopping coefficients do not change the position of the global minima in the single-particle spectrum but generate metastable states. 2D separable potentials do not introduce new physics from this point of view.
The simplest non-separable potential in 2D is given by \cite{hemmerich2}
\begin{align}
\begin{split}
V(\b{r}) = -V_0\Bigl\{ & \, \sin^2[k(x+y)] + \sin^2[k(x-y)] \\
& \, + 2\alpha\sin[k(x+y)]\sin[k(x-y)]\Bigr\}, 
\label{nonsep}
\end{split}
\end{align}
where $k = 2 \sqrt{2} \pi / 2 \lambda$ (the factor $\sqrt{2}/2$ comes from a coordinates transformation corresponding to a rotation of the lattice of $\pi/4$) and we will make the choice $\alpha=1$ in the remainder of this work. Had we chosen $\alpha = 0$, the potential would have been separable, whereas for $0 < \alpha < 1$ the potential would correspond to a superlattice, with neighboring wells of different depths.
\begin{figure}[!htbp]	
	\begin{center}
	\includegraphics[width=0.35\textwidth]{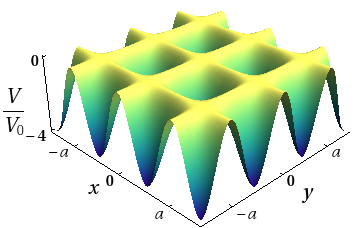}
	\end{center}
	\caption{Non-separable optical potential $V(x, y)$ given by Eq.\ (\ref{nonsep}) with 
	$\alpha=1$.}
	\label{NSpot}
\end{figure}

As shown in e.g.~Ref.\,\cite{bloch}, we can calculate the hopping coefficients from the exact band structure
\be 
E_n(\b{q}) = \sum_{\b{R}} t_n(\b{R})\,e^{i\b{q}\cdot\b{R}}\,,
\ee
where $n$ is the band index, $q$ is the quasimomentum, and $\b{R}$ is a lattice vector. In this notation, $t_n(\b{R})$ is the hopping coefficient between two sites separated by the lattice vector $\b{R}$ in the $n$-th energy band. The non-separable optical potential generates hopping coefficients along directions other than those of the elementary lattice vectors of the lattice which were exactly zero for separable potentials. A lattice vector has the form $\b{R}=m a\b{e}_x + n a \b{e}_y$, where $a = \lambda / \sqrt{2}$ is the lattice spacing, $m$ and $n$ are integers, and $\b{e}_x$ and $\b{e}_y$ are unit vectors in the $x$- and $y$-directions; $\b{R}$ is indicated in short notation as $(m,n)$. For the non-separable potential that we have introduced, we find non zero hopping terms also for pairs of sites identified by (1,1) or (2,1), which vanish for separable lattices. Table\,\ref{T2} shows the most relevant lowest-band hopping coefficients for shallow lattices. Higher-order hopping coefficients are neglected because they are at least ten times smaller than $t''$ and therefore not important, as will become clear afterwards.
 
\begin{table}[!htbp]
\begin{center}
\begin{tabular}{c c c c c c c} \hline\hline
$ V_0/E_{r}$  &  $(1,0)\lra -t$  &  $(1,1)\lra t'$     & $(2,0)\lra t''$   \\ \hline
$ 1.0$ & $-2.45\times10^{-2}$ & $-8.89\times10^{-4}$ & $8.88\times10^{-4}$ \\
$ 2.0$ & $-4.52\times10^{-3}$ & $-6.65\times10^{-5}$ & $2.27\times10^{-5}$  \\
$ 3.0$ & $-1.06\times10^{-3}$ & $-5.89\times10^{-6}$ & $1.06\times10^{-6}$ \\
$ 4.0$ & $-2.97\times10^{-4}$ & $-6.74\times10^{-7}$ & $7.86\times10^{-8}$  \\
\hline\hline
\end{tabular}
\caption{Relevant hopping matrix elements (in units of the recoil energy $E_r$) of the lowest band for shallow lattices.}
\label{T2}
\end{center}
\end{table}
We will assume that the lowest-orbital Wannier functions are still even and real for this non-separable potential. As shown by Kohn \cite{kohn}, this can be proven for separable potentials; for non-separable ones it is also a reasonable conjecture, supported by numerical simulations, as shown in Ref.\ \cite{marzari}. If we apply a driving sinusoidal force like the one studied in Ref.\,\cite{eckardt}, but now along one of the diagonals, the shaking term in the co-moving reference frame that has to be added to the Hamiltonian reads 
\be
W(\tau)=K\cos(\om \tau)\sum_{i,j}(i+j)n_{ij}\,,
\ee
where $\om$ is the shaking frequency, $\tau$ is the real time, and $n_{ij}$ is the density operator at site $(i,j)$. Following the approach discussed in Refs.\,\cite{eckardt,hemmerich}, the non-interacting effective hamiltonian for the quasienergy spectrum in the high-frequency limit $\hbar\om\gg U,t$ (and thus $\hbar\om\gg t',t''$) is
\begin{widetext}
 \bea 
\label{2Deff}
H^0_{\rm{eff}} = -t J_0(K_0) \sum_{\b{r},\nu=x,y}a^\dag_{\b{r}}a_{\b{r}\pm\b{e}_\nu} + t' J_0(2K_0) \sum_{\b{r}}
a^\dag_{\b{r}}a_{\b{r}\pm(\b{e}_x+\b{e}_y)}+
t'\sum_{\b{r}}a^\dag_{\b{r}}a_{\b{r}\pm(\b{e}_x-\b{e}_y)} +t'' J_0(2K_0) \sum_{\b{r},\nu=x,y}a^\dag_{\b{r}}a_{\b{r}\pm2\b{e}_\nu}, 
\eea
\end{widetext}
where the shaking parameter is $K_0=K/\hbar\om$. The Bessel function $J_0(x)$ has a node at $x\simeq 2.4048$; hence, when the nn-hopping coefficient $t_{\rm{eff}}= t\, J_0(K_0)$ is negligible, the  higher-order ones are not.  Note that the hopping coefficient along the diagonal perpendicular to the shaking direction is not affected by the shaking. 

\section{TUNABLE FINITE-MOMENTUM CONDENSATE}

\begin{figure}[!htbp]	
	\begin{center}
	\includegraphics[angle=-2,width=0.3\textwidth]{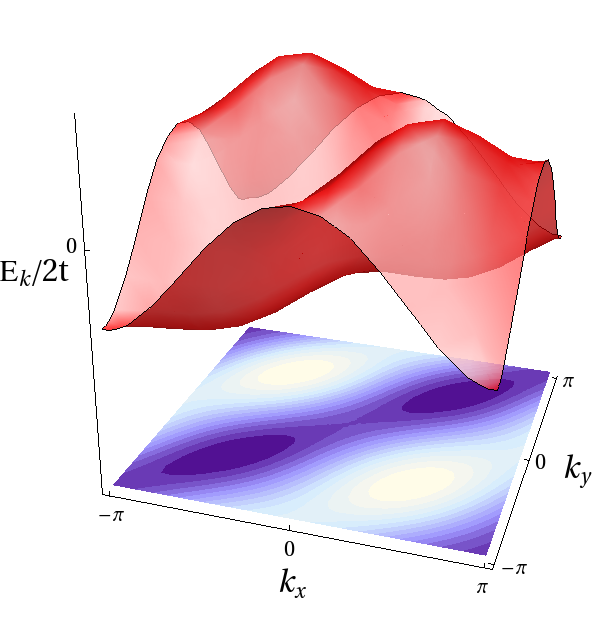}
	\end{center}
	\caption{Single-particle spectrum at $V_0 =3E_r$ for $K_0=2.4048$ and contour plot.}
	\label{finmom}
\end{figure}

\begin{figure}[tbh]	
	\begin{center}
	\includegraphics[width=0.35\textwidth]{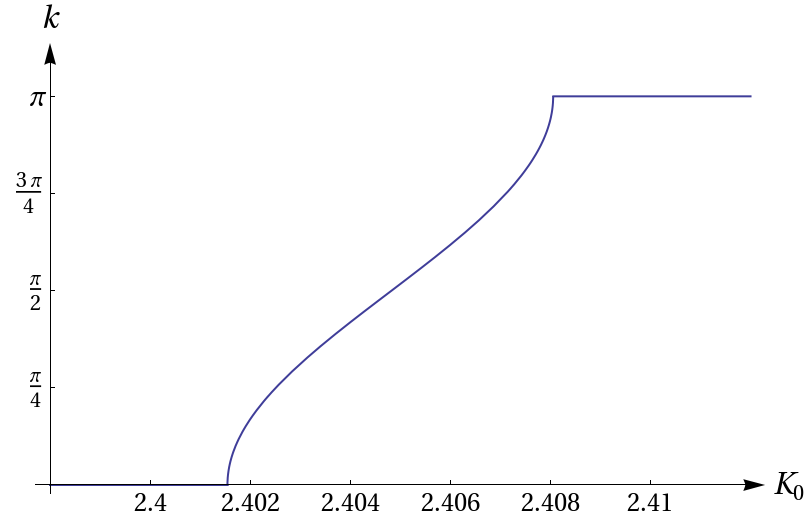}
	\end{center}
	\caption{Evolution of the minimum in the single-particle spectrum in units of the lattice spacing $a$ as a function of the 
shaking parameter $K_0$ at $V_0 =3E_r$ (only the positive branch is considered).}
	\label{momentum}
\end{figure}

The effective Hamiltonian is diagonal in reciprocal space and the single-particle spectrum reads
\bea
\label{spectrum}
E_\b{k} &=& -2tJ_0(K_0)\left[\cos(k_x )+\cos(k_y )\right]   \nn\\
&& +2t'J_0(2K_0)\cos(k_x+k_y) +2t'\cos(k_x-k_y)  \nn\\
&& +2t''J_0(2K_0)\left[\cos(2k_x )+\cos(2k_y )\right]\,,
\eea
where $k_\nu = \b{k}\cdot\b{e}_\nu$ and we have set the lattice constant to unity. The spectrum has an absolute minimum at the center of the Brillouin zone ($\b{k}=0$) when $K_0 < 2.4048 - \delta$ and at the four corners of the Brillouin zone when $K_0 > 2.4048 + \delta$. In the interval $2.4048 - \delta < K_0 < 2.4048 + \delta$, two symmetric minima develop along one diagonal of the Brillouin zone at $\pm\b{k}_0$, as shown in Fig.\,\ref{finmom}. 
We may determine the precise position of these two minima by studying the first derivative of the single-particle spectrum for $k=k_x=k_y$. The non-trivial minima are given by the solution of the equation 
\be
\label{fx}
\cos(ka)=\f{J_0(K_0)}{2J_0(2K_0)(t_1+2t_2)}\equiv f(K_0)\,.
\ee
where $t_1=t'/t$ and $t_2=t''/t$. We have found that for $V_0=2E_r,3E_r,$ and $4E_r$, the second derivative of the single-particle spectrum shows that Eq.(\ref{fx}) corresponds to a true minimum, while for $V_0=1E_r$ it is a maximum. The largest interval $\Sigma = 2 \delta$ of the shaking parameter $K_0$ for which the non-trivial minima appear has been found to be at lattice depth $V_0=2.2E_r$, where $\delta = 0.0045$, and hence the condensation momentum is finite for $2.4003 < K_0 < 2.4093$. 
Since we expect the bosons to condense at the minimum of the single-particle spectrum, the condensation momentum given by Eq.\,(\ref{fx}) is a function of the shaking parameter $K_0$ and smoothly evolves from $\b k = \b{0}$ at the left edge of $\Sigma$ to $\b k = (\pi,\pi)$ at the right edge of $\Sigma$, see Fig.\ \ref{momentum}. The two minima in the $\Sigma$ region are inequivalent because they are not connected by reciprocal lattice vectors and we thus need to take both into account for evaluating the condensation momentum. The arccosine shape of the evolution curve can be explained by linearising Eq.~\eqref{fx} around $K_0 = 2.4048$, which is a good approximation because $\delta \ll 1$.

The non-interacting ground state of the tunable-momentum SF phase with momenta $\pm\b{k}_0$ is
\bea
\label{g1}
|G\ket &=& \sum_{n=0}^{N}\f{c_n}{\sqrt{n!(N-n)!}}(a^\dag_{\b{k}_0})^{n}(a^\dag_{-\b{k}_0})^{N-n}|0\ket\nn\\
&=& \sum_{n=0}^{N}c_n|n_{\b{k}_0},(N-n)_{-\b{k}_0}\ket\,,
\eea
where the coefficients $c_n$ obey the normalization condition $\sum_n |c_n|^2=1$. The ground state is thus $N+1$-fold degenerate, where $N$ is the number of particles. 

We stress that there is a close similarity between our system and a BEC of magnons (or triplons). In dimerized antiferromagnets, the magnons condense at a non-zero wavevector $\b{k}_0=(\pi/a,\pi/a)$ for applied magnetic fields $H$ which lie between two critical values, for $H_{c1}< H < H_{c2}$  \cite{giamarchi}. In addition, the magnon dispersion of a two-leg antiferromagnetic ladder with frustrated nnn couplings along the legs shows a minimum that is incommensurate with the lattice spacing \cite{tsirlin}.

\section{PHASE DIAGRAM}

Let us now consider an additional term to the Hamiltonian (\ref{2Deff}), which describes the local interactions between the bosons
\be
H_{\textrm{int}}=\frac U 2\sum_{\b{r}}n_{\b{r}}(n_{\b{r}}-1).
\ee 
We will treat the interactions between the atoms in a perturbative way and study their effect on the ground state degeneracy. By applying first-order perturbation theory, we find that the correction to the ground-state energy $N E_{\b{k}_0}$ is 
given by
\bea
\lefteqn{\bra m,N-m| H_{\rm{int}}|n,N-n\ket =}\nn\\
&&= \f{U}{2N_s} \left[-2n^2+2nN+N(N-1)\right] \delta_{mn}\,,
\eea
where $N_s$ is the number of lattice sites. The matrix element is in diagonal form and the eigenvalues are an upside down parabola in $n$. This means that the minima are at the edge of the interval $n\in[0,N]$ and that they are degenerate. Interactions have partially removed the  degeneracy; the perturbative (degenerate) ground state to zeroth order is 
\be
\label{g'}
|G\ket = \f{c_+}{\sqrt{N!}}(a^\dag_{\b{k}_0})^{N}|0\ket + \f{c_-}{\sqrt{N!}}(a^\dag_{-\b{k}_0})^{N}|0\ket
\ee
and has energy
\be
\bra H \ket=\bra H_0 \ket+\bra H_{\rm{int}} \ket=N E_{\b{k}_0}+\f{U}{2N_s} N (N-1). 
\ee
Eq.\ (\ref{g'}) shows that the ground state is a superposition of two degenerate states in which all the particles have momentum $\b{k}_0$ or $-\b{k}_0$. These two states are entangled and behave in  a very similar way to the states found by Stanescu \emph{et al.} \cite{stanescu} for condensates with spin-orbit coupling. 

\begin{figure}[!tbp]	
	\begin{center}
	\includegraphics[width=0.40\textwidth]{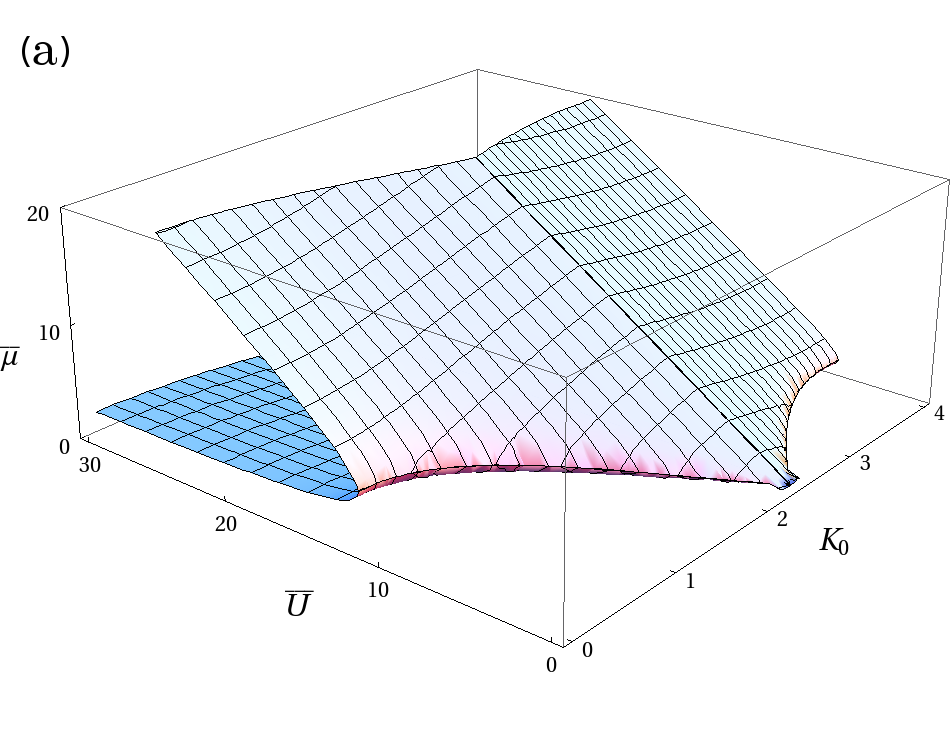}
	\includegraphics[width=0.40\textwidth]{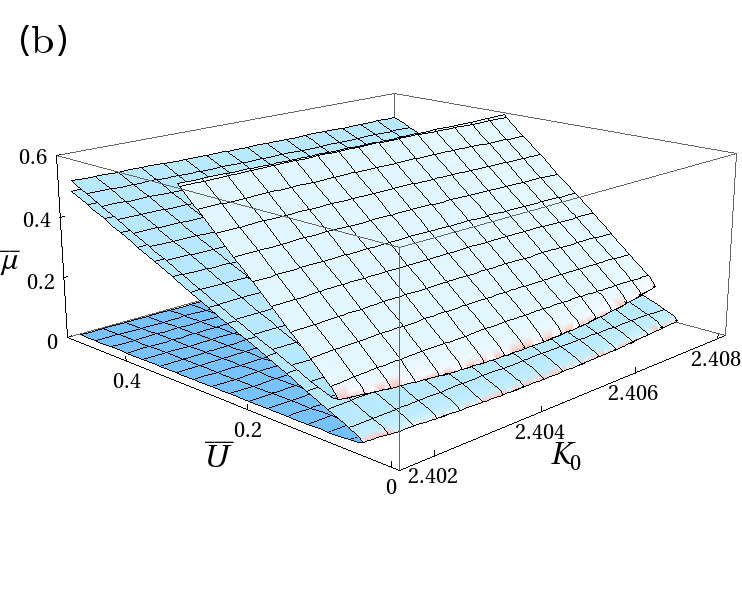}
	\end{center}
	\caption{Phase boundaries for $V_0=3 E_r$ where $\bar{\mu}\equiv\mu/2t$ and $\bar{U}\equiv U/2t$: (a) lobe with $N_0=1$; (b) lobes with $N_0=1,2$ in the region of the tunable finite-momentum condensate.}
	\label{Lobes1}
\end{figure}

\begin{figure}[!tbp]	
	\begin{center}
	\includegraphics[width=0.30\textwidth]{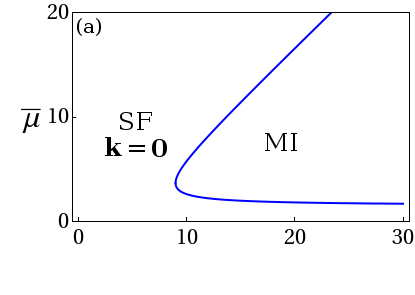}
	\includegraphics[width=0.30\textwidth]{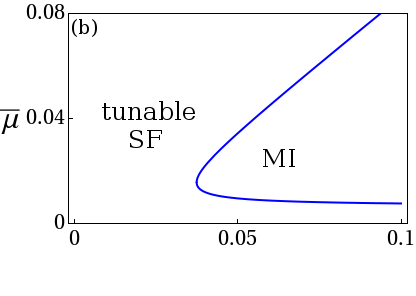}
	\includegraphics[width=0.30\textwidth]{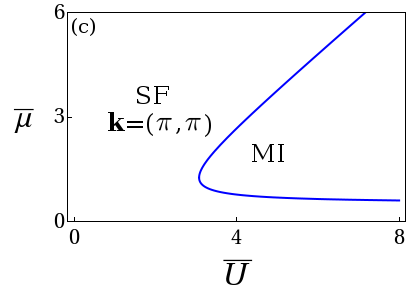}
	\end{center}
	\caption{Phase boundaries between the Mott-insulator and the superfluid phase for $V_0=3 E_r$ at fixed $K_0$ and filling factor $N_0=1$: (a) $K_0 = 1$; (b) $K_0=2.405$; (c) $K_0=3$. Note the different scales for each plot.}
	\label{Lobes2}
\end{figure}

One can generalize the approach described in Ref.\,\cite{dries} to calculate the MI-SF phase boundaries, taking into account higher-order hopping terms. 
The outcome is 
\bea
\lefteqn{\bar{\mu}_\pm=\f{\bar{U}}{2}(2N_0-1)+}\nn\\
&&+\f{\eps_{\b{k}_0}}{2}\pm\f1 2\sqrt{\eps^2_{\b{k}_0}+2(2N_0+1)\bar{U}\eps_{\b{k}_0} + \bar{U}^2}\,,
\eea
where $\bar{\mu}=\mu/2t$, $\bar{U}=U/2t$, $\eps_{{\b{k}_0}}=E_{{\b{k}_0}}/2t$ and $\b{k}_0$ is the condensation momentum, which depends on the shaking parameter $K_0$. Plotting $\bar{\mu}_\pm$ then gives the phase diagram, which is shown in Figs.\,\ref{Lobes1} and \ref{Lobes2}. We note that the condensation momentum is not changed by the interactions. This can be seen e.g.~by doing first order perturbation theory calculations: in the presence of interactions, the energy per particle is shifted by an amount of $NU/2N_s$, which is momentum independent. 

\section{Experimental considerations}

\begin{figure}[!tbp]	
	\begin{center}
	\includegraphics[width=0.4\textwidth]{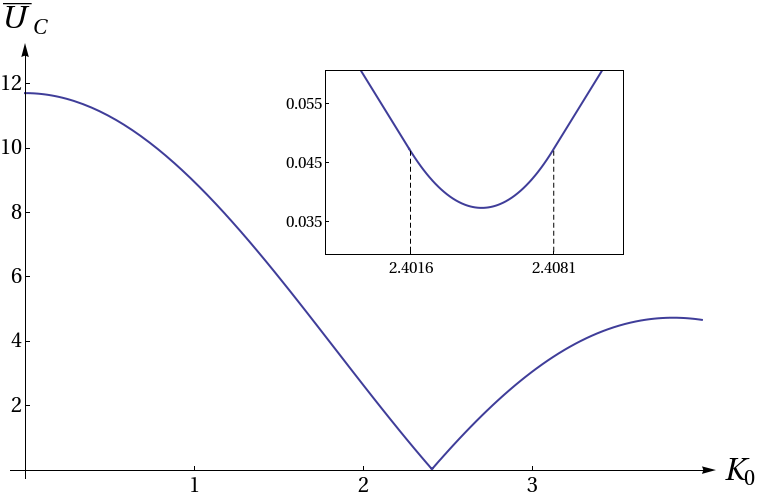}
	\end{center}
	\caption{Critical value $\bar{U}_c=(U/2t)_c$ at $V_0 =3E_r$ as a
	function of the shaking parameter $K_0$; in the inset, we show the region $2.4016\leq K_0 \leq 2.4081$ where the tunable finite-momentum condensate is generated.}
	\label{PBV3}
\end{figure}

The lobe with unit filling $N_0=1$ yields a critical value of $\bar{U}$ below which only the SF phase is allowed, see Fig.\,\ref{PBV3}. Typical values of $U$ are too large to allow us to probe the tunable-momentum SF with ordinary experimental setups. However, we can decrease $U$ by reducing the $s$-wave scattering length with Feshbach resonances, which are available for both the Rubidium isotopes and also for other alkali atoms.
We remark that although the range of $K_0$ in which the condensation momentum is tunable is very small, the required precision is well within experimental control \cite{arimondo}.

\begin{figure}[!tbp]	
	\begin{center}
	\includegraphics[width=0.4\textwidth]{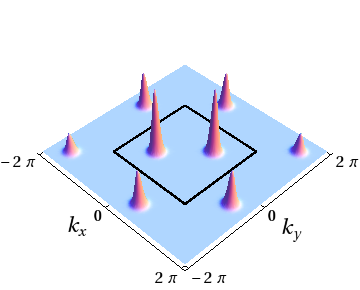}
	\end{center}
	\caption{Time-of-flight picture expected from experiment as a signal of the finite-momentum-condensate phase. The black square indicates the first Brillouin zone, and the condensation momentum represented in this image is $\b{k}_0 = {2 \pi / 5 , 2 \pi / 5 }$. }
	\label{TOF}
\end{figure}

The quantum phases discussed above could be experimentally observed by doing the usual time-of-flight experiments. These experiments measure the momentum-space density distribution 
\bea
\label{eq:tof}
n(\b{k})&=&\bra\psi^\dag(\b{k})\psi(\b{k})\ket\nn\\
&=&N|W(\b{k})|^2\left(|c_+|^2\delta_{\b{k},\b{k}_0}+|c_-|^2\delta_{\b{k},-\b{k}_0}\right),
\eea
where $W(\b{k})$ is the Fourier transform of the Wannier function and we adopted the coherent state approximation for the SF ground state. The delta functions select the positions of the peaks in the absorption image and are a clear signal of the presence of such a condensate. When the image is recorded, and the first atom is measured to have one of the two momenta, then the wave function collapses on that state, showing only one peak. An array of identical 2D systems would reveal a pattern with both peaks and we can study the effect of the non separability of the optical potential on the Wannier functions in reciprocal space. In Fig.\ref{TOF}, we show a qualitative indication of the time-of-flight image described by Eq.(\ref{eq:tof}). We have lumped together the effects of a hypothetical external trap and the Fourier transform of the Wannier function into a Gaussian filter, suppressing the peak heights in higher Brillouin zones. In addition, we have modeled the broadening of the peaks by replacing every peak by a highly localised Gaussian. 

It is instructive to compare this pattern with predictions for other systems like the time-of-flight images given in e.g.~Refs.~\cite{lim2, olivier} for finite-momentum superfluids and supersolids. We clearly see that these phases have a pattern different from Fig.\ref{TOF}, because the position of the peaks differs. Hence, we can be sure that we would have an unambigous signal of the measurement of the tunable-momentum superfluid phase from experiments.

\section{DISCUSSIONS AND CONCLUSIONS}

In conclusion, we have explored the possibility to generate finite-momentum condensates in optical lattices under shaking, where the suppression of hopping can be tuned by the shaking. This opens up the possibility to investigate the role of higher-order hopping. To look for nontrivial condensation points lying inside the first Brillouin zone, we have studied non-separable optical potentials in 2D square lattices. By applying the shaking along the diagonal of the lattice, we found that  in a small region of the shaking parameter, where the nn tunneling is suppressed, two kinds of higher-order-hopping coefficients govern the dynamics of the condensate. In this region, we have unveiled an intermediate phase, for which the condensation point varies continuously from the center to the edge of the Brillouin zone as we tune the shaking parameter. There are two minima in the single-particle spectrum and they are symmetric with respect to the center of the Brillouin zone.
In addition, we found that small interactions between the particles force the ground state to be a superposition of two possible Fock states: one where all the particles condense in one minimum and the other where all the particles condense in the second minimum.

Finally, we note that the tunable-momentum condensate can be measured experimentally if the on-site interaction is reduced significantly. This can be achieved with present state-of-the-art experimental techniques, and we hope that our results can stimulate further experiments in this direction. 
This work yields new insights for the realization and control of finite-momentum condensates and opens new possibilities in the search for bosonic analogs of the FFLO superconducting phase. 

\section*{ACKNOWLEDGMENTS}
It is a pleasure to acknowledge A. Hemmerich, E. Arimondo, D. Makogon, A. Eckardt, N. Goldman, and I. Spielman for fruitful discussions. This work was partially supported by the Netherlands Organization for Scientific Research (NWO) and the Scuola Superiore di Catania. 

\addcontentsline{toc}{chapter}{Bibliography}
\bibliographystyle{mprsty}
\bibliography{Biblio}

\end{document}